# -55°C TO 170°C HIGH LINEAR VOLTAGE REFERENCES CIRCUITRY IN 0.18μm CMOS TECHNOLOGY

*Joseph Tzuo-sheng Tsai and Herming Chiueh*

Nanoelectronics and Infotronic Systems Research Center
and Department of Communication Engineering,
National Chiao Tung University, Hsinchu 300, TAIWAN
Tel:+886-3-5712121 ext 54597 Fax: +886-3-5710116
Jorsan.cm93g@nctu.edu.tw, chiueh@ieee.org

## ABSTRACT

High linear voltage references circuitry are designed and implemented in TSMC 0.18μm CMOS technology. Previous research has proposed the use of MOS transistors operating in the weak inversion region to replace the bipolar devices in conventional PTAT(proportional to absolute temperature) circuits. However such solutions often have linearity problem in high temperature region due to the current leaking devices in modern deep sub micron and nano-scale CMOS technology. The proposed circuit utilized temperature complementation technique on two voltage references, PTAT and IOAT (independent of absolute temperature) references, to enhance the linearity and produce a more stable IOAT voltage reference. Base on the simulation results, the R-squares of both circuitries are better than 0.999 in a considerable wider temperature range from -55°C to 170°C. Thus, a fully integrated temperature sensor with wider temperature range is designed and easily to integrate to modern system-on-chip designs with minimal efforts.

## 1. INTRODUCTION

Increases in circuit density and clock speed in modern VLSI systems have brought thermal issues into the spotlight of high-speed VLSI design. Large gate-count and high operating frequency in modern system-on-chip integration escalate the problem. Previous research has indicated that the thermal problem can cause significant performance decay [1] as well as reducing of circuitry reliability [2]-[5]. In order to avoid thermal damages, early detection of overheating and properly handling such event are necessary. For these reasons, temperature sensors are widely used in modern VLSI systems.

Recent research has indicated that the best candidate for a fully-integrated temperature sensor is the proportional-to-absolute temperature (PTAT) circuit [6] and IOAT circuit with the sigma-delta modulator and digital filter. In such design, the PTAT sources are usually implemented using parasitic vertical BJTs in any standard CMOS technology [7], [8]. These circuits require resistors which may vary from different technology. Also, the power consumption of the BJT based references is relatively high for low power applications. However, in deep sub micron CMOS technology, the characteristic of vertical BJT is getting worse. So, the design of temperature sensor has become a major challenge in deep sub micron technology.

The PTAT generator of Vittoz and Fellrath [9] takes advantage of MOS transistors operating in the weak inversion region; the power consumption is minimal due to the inherently low currents in that region. However, this circuit does not allow strong supply voltage scaling. Serra-Graells and Huertas [10] introduce an all-MOS implementation exhibiting enough low-voltage capabilities by the use of MOS sub-threshold techniques. However, in this circuit, the current leaking device in modern deep sub-micron CMOS technology has cause the linearity problem of the PTAT and IOAT signals in high temperature range.

These nonlinearity behaviors are crucial effect to implement a complete thermal management system within a digital circuit since such circuitries require more efforts and costs for after process calibration. Thus, linearity and power issues are the key factors for design a fully integrated temperature sensor in the deep sub micron CMOS technology.

In this paper, both PTAT and IOAT voltage references are redesigned by utilizing sub-threshold MOSFETs and temperature complementation technique to enhance the linearity and produce a more stable output. The propose design has extend the linear temperature rage of on-chip temperature sensor to -55°C to 170°C which provides a practical solution for modern system-on-chip's thermal management systems. The design concept of proposed circuit will be described in Section 2. Following in Section 3, simulation result is presented. Finally, a conclusion is summarized in Section 4.

## 2. CIRCUITRY DESIGN

In this section, the gate-source voltage formula operated in weak inversion is verified in different technologies. Base on the verification, two new voltage generator circuitries is designed (PTAT and IOAT). In order to generate two voltage references, PTAT is designed design firstly, and the IOAT voltage reference is generated by using PTAT reference as one of the inputs.

Previous research [11] has shown the gate-source voltage of an nMOS which operated in weak inversion has a negative temperature coefficient (nTC) and can been modeled as:

$$V_{GSn(T)} \approx V_{GSn}(T_0) + K_{Gn} \cdot \left(\frac{T}{T_0} - 1\right) \qquad (1)$$

where

$$K_{Gn} \cong K_{Tn} + V_{GSn}(T_0) - V_{THn}(T_0) - V_{OFFn} \qquad (2)$$

The gate-source voltage of a pMOS transistor can also been modeled as

$$\left|V_{GSp}(T)\right| \approx \left|V_{GSp}(T_0)\right| + K_{Gp} \cdot \left(\frac{T}{T_0} - 1\right) \qquad (3)$$

where

$$K_{Gp} \cong K_{Tp} + \left|V_{GSp}(T_0)\right| - \left|V_{THp}(T_0)\right| - V_{OFFp} \qquad (4)$$

In order to verify the derived model in deep sub micron simulations based on both TSMC 0.18µm and 0.35µm technology are conduced, the gate-source voltage of an nMOS diode-connected transistor biased with a 100-nA current and the diode aspect ratio was set to 50/2 are simulated. The same simulations are also done with a pMOS diode-connected transistor. The results are shown in Fig.2 and Fig.3. Basing on the results shown in Fig. 1 and Fig. 2, we can know that the error percentage of nMOS's model is smaller than the pMOS's in TSMC 0.18µm CMOS technology. In TSMC 0.35µm CMOS technology, the result is opposite. The simulation shows that the wide temperature range linearity in TSMC 0.18µm CMOS technology is better than in TSMC 0.35µm CMOS technology. So, if we want to get more better linearity in wider range, for the case, [-55, +170]℃, the gate-source voltage model of an nMOS transistor is the best choice.

Above all, we know the gate-source voltage of an nMOS transistor operated in the weak inversion region has a linear negative temperature coefficient (nTC) and is suit for our design. So if we put PTAT core and VGSn (weak inversion) together, the IOAT voltage reference will be achieved by sum up both out. According to previous researches [8], a PTAT voltage reference circuit based on subthreshold MOSFETs has been developed. Fig. 3 illustrates the condensed scheme of two low-voltage CMOS PTAT references [12]. M1, M2, and compensation transistor Mc operate in weak inversion region while transistors M3-M8 ensure the current ratio of M1-M2 pair.

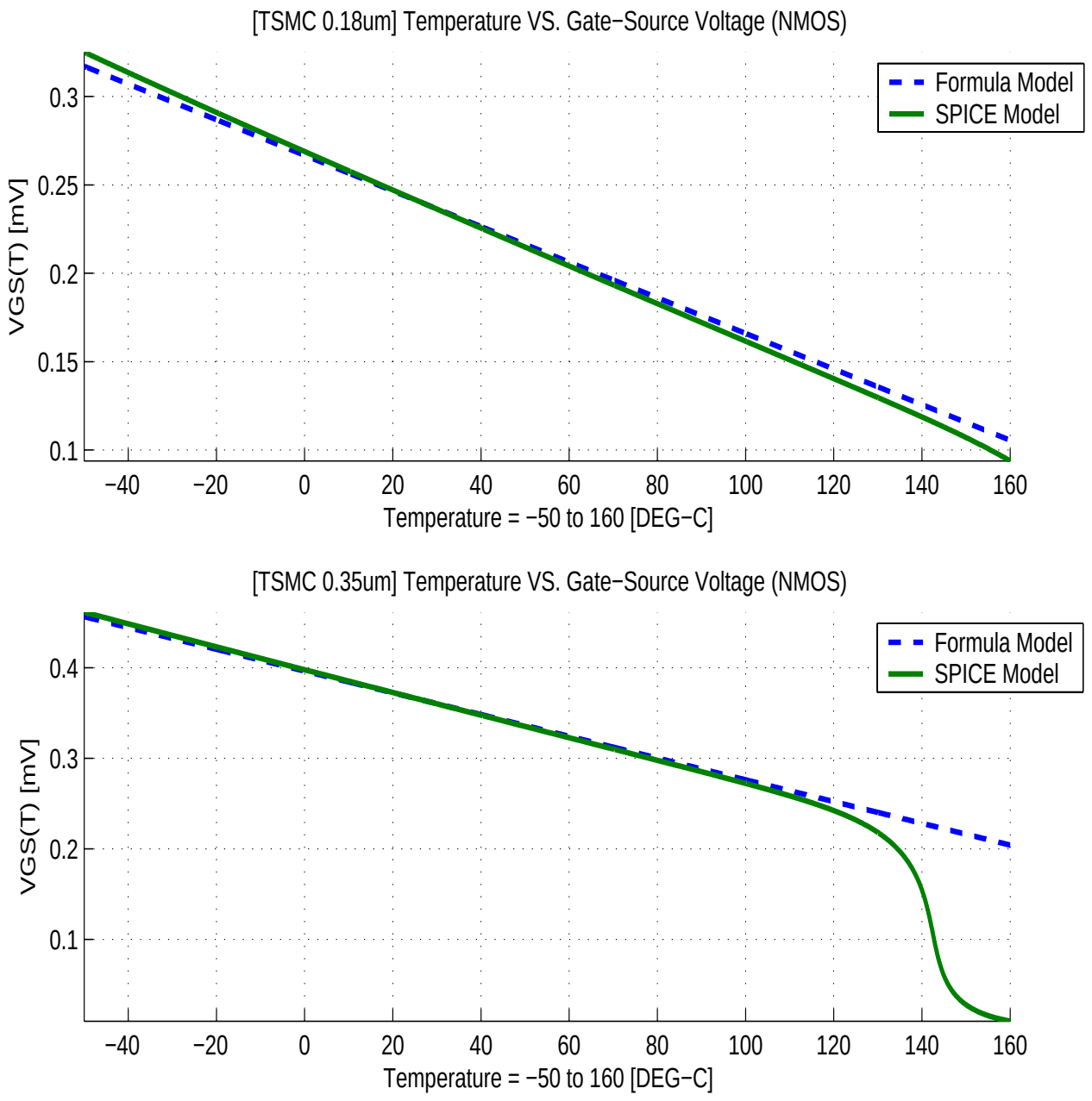

Fig. 1. The simulation result of an nMOS diode-connected transistor.

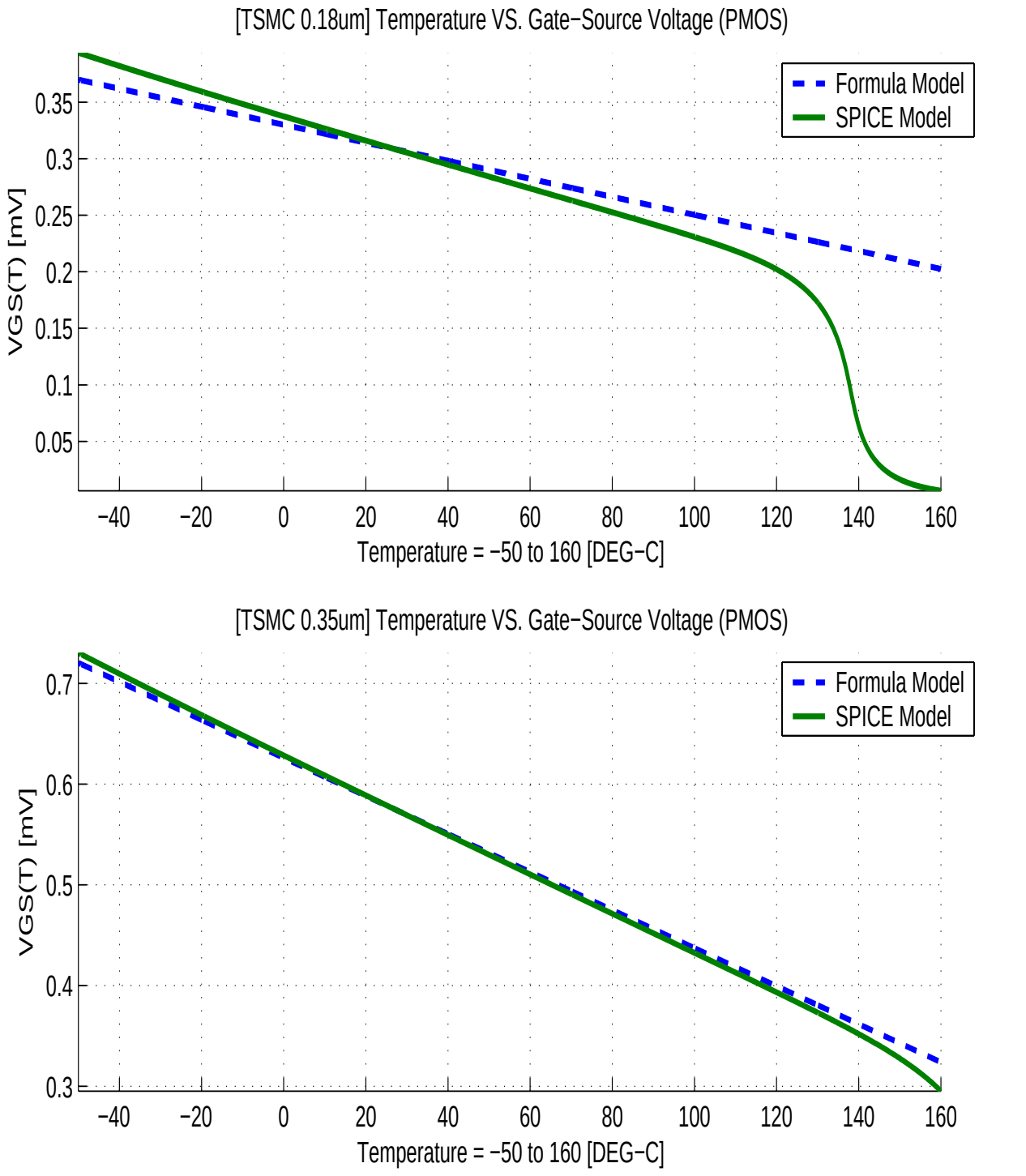

Fig. 2. The simulation result of a pMOS diode-connected transistor.

The proposed circuitry architectures are shown in Fig. 4 and Fig. 5. The design concept is that using current mirror combines positive and negative temperature coefficients. Fig. 5 shows resistor-based PTAT and IOAT voltage references. The first part circuit, M1-M8, Mc, and R1, will produce a PTAT voltage reference. The second part of this circuit is made up of M9, R2, and a diode-connected transistor, Mn. A negative temperature coefficient will be produced in the gate-source voltage of Mn. The target of our design is to make two different temperature coefficient sum up, so we use a current mirror to make them sun up in current type. In this architecture, the VIOAT can be expressed as

$$V_{IOAT} = V_{PTAT} \cdot \frac{R_2}{R_1} \cdot \frac{S_9}{S_8} + V_{GSn} \quad (5)$$

S8 and S9 are the aspect ratios of M8 and M9.

For the area consideration, we also develop all-MOS PTAT and IOAT voltage reference. Fig. 6 shows the all-MOS architecture. Following (6), (7), (8), (9)

$$I_D = \beta \cdot [V_{GB} - V_{TO} - \frac{n}{2}(V_{DB} + V_{SB})] \cdot (V_{DB} - V_{SB}) \quad \text{s.i.cond.} \quad (6)$$

$$I_D = \frac{\beta}{2n} \cdot (V_{GB} - V_{TO} - nV_{SB})^2 \quad \text{s.i.sat.} \quad (7)$$

$$I_D = I_S \cdot e^{(V_{GB}-V_{TO})/nU_t} \cdot e^{-V_{SB}/U_t} \quad \text{w.i.sat.} \quad (8)$$

$$I_S = 2 \cdot n \cdot \beta \cdot U_t^2 \quad (9)$$

where VTO, β, n, and Is stand for the threshold voltage, current factor, subthreshold slope, and specific current, respectively, as defined in the EKV model [13], we can get the VIOAT as

$$V_{IOAT} = V_{GSn13} + k \cdot V_{PTAT} \quad (10)$$

where

$$k = \frac{Q+1}{M+1} \cdot \frac{1 + \sqrt{1 + N + \frac{N}{M}}}{1 + \sqrt{1 + S + \frac{S}{Q}}} \cdot \sqrt{\frac{M}{N} \cdot \frac{S}{Q}} \quad (11)$$

In this section, we proposed two circuitry architectures, resistor-based and all MOS voltage generators.

## 3. EXPERIMENTAL RESULT

In Section 2, we proposed two new circuitry architectures, resistor-based and all MOS voltage generators and have been complete described. In this section, two PTAT and IOAT references including resistor-based and all-MOS references have been simulated in TSMC 0.18µm 1P6M standard CMOS technology.

Fig. 6 shows the PTAT voltage versus temperature for all-MOS and resistor-based PTAT references. The simulation range is from -55°C to 170°C temperature range is conduced for each circuit. The R-squares of resistor-based and all-MOS circuits are 0.99963 and 0.99968 respectively. The temperature behavior of resistor-based generator is similar to the all-MOS implementation.

Fig. 7 shows the PTAT voltage versus temperature for all MOS IOAT references. The simulation range is from -55°C to 170°C for each circuit. The means of resistor-based and all-MOS circuits are 578.75mV and 514.94mV respectively. The variation of the resistor-based circuit is about ±5mV. For the all-MOS circuits, the variation is about ±8mV. The simulation results show that the linearity of all-MOS circuits is better. For variation of IOAT reference, however, resistor-based circuitry is better than all-MOS. In Fig. 8, we put Fig. 6 and Fig.7 together. It summarizes the simulation results. For the PTAT linearity, all MOS circuitry is better than resistor-based circuitry. For the IOAT variation, resistor-based circuitry is better than all MOS circuitry.

## 4. CONCLUSION

In this paper, -55°C to 170°C high linear voltage references circuitry for fully integrated temperature sensor is designed and implemented in TSMC 0.18µm CMOS technology. The proposed circuit utilized temperature complementation technique on PTAT and IOAT references. Base on the simulation results, the R-squares of both circuitries are better than 0.999. in a considerable wider temperature range from -55°C to 170°C as shown in Table 1. Thus, a fully integrated temperature sensor with wider temperature range is designed and easily to integrate to modern system-on-chip designs with minimal efforts.

Table I. Summary of simulation results

| | PTAT Temperature Coefficient (mV / °C) | PTAT R-square | IOAT Mean (mV) |
|---|---|---|---|
| R-based | 0.206 | 0.99963 | 514.94 |
| All-MOS | 0.276 | 0.99968 | 578.75 |

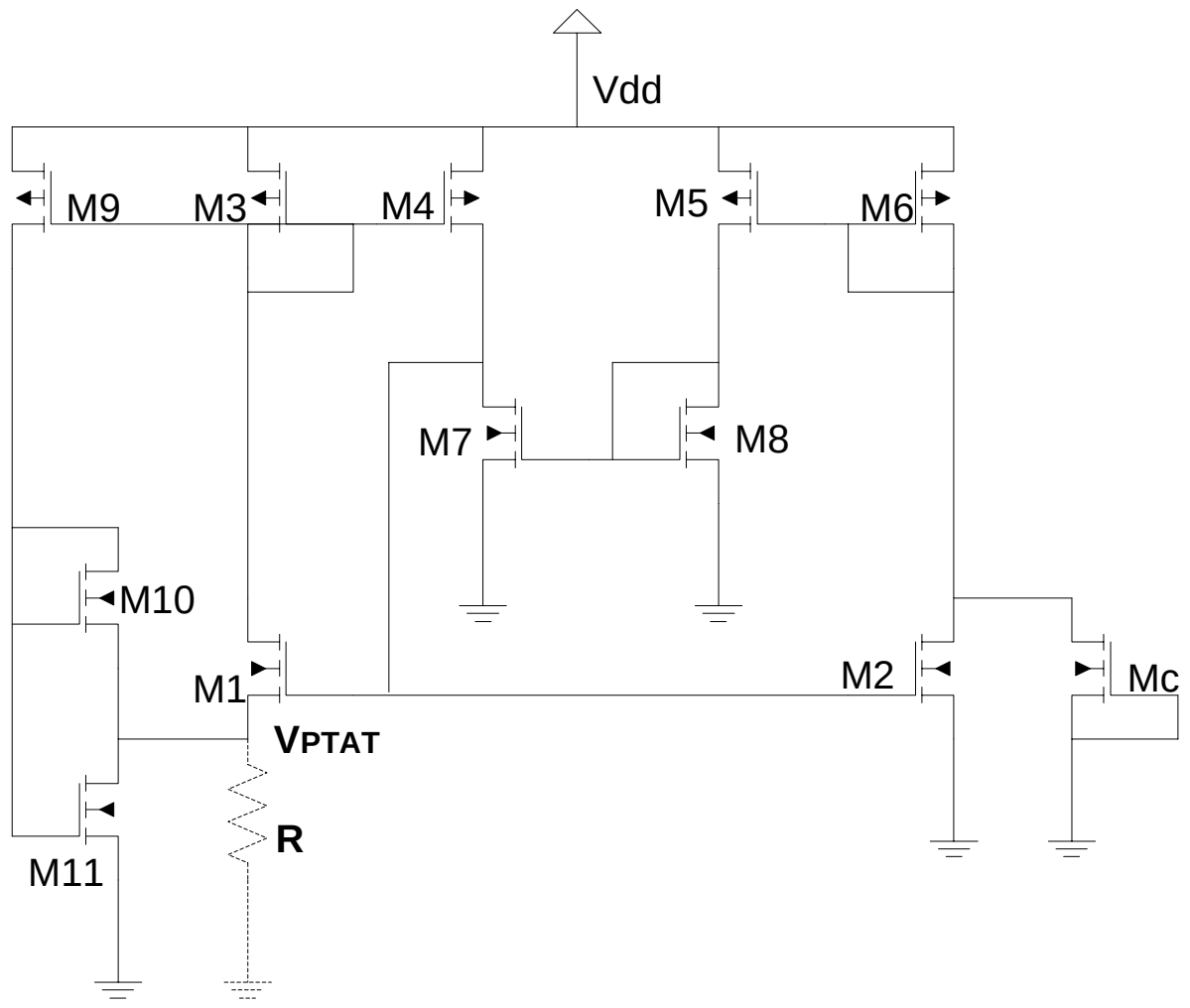

Fig. 3. Low-voltage CMOS PTAT references.

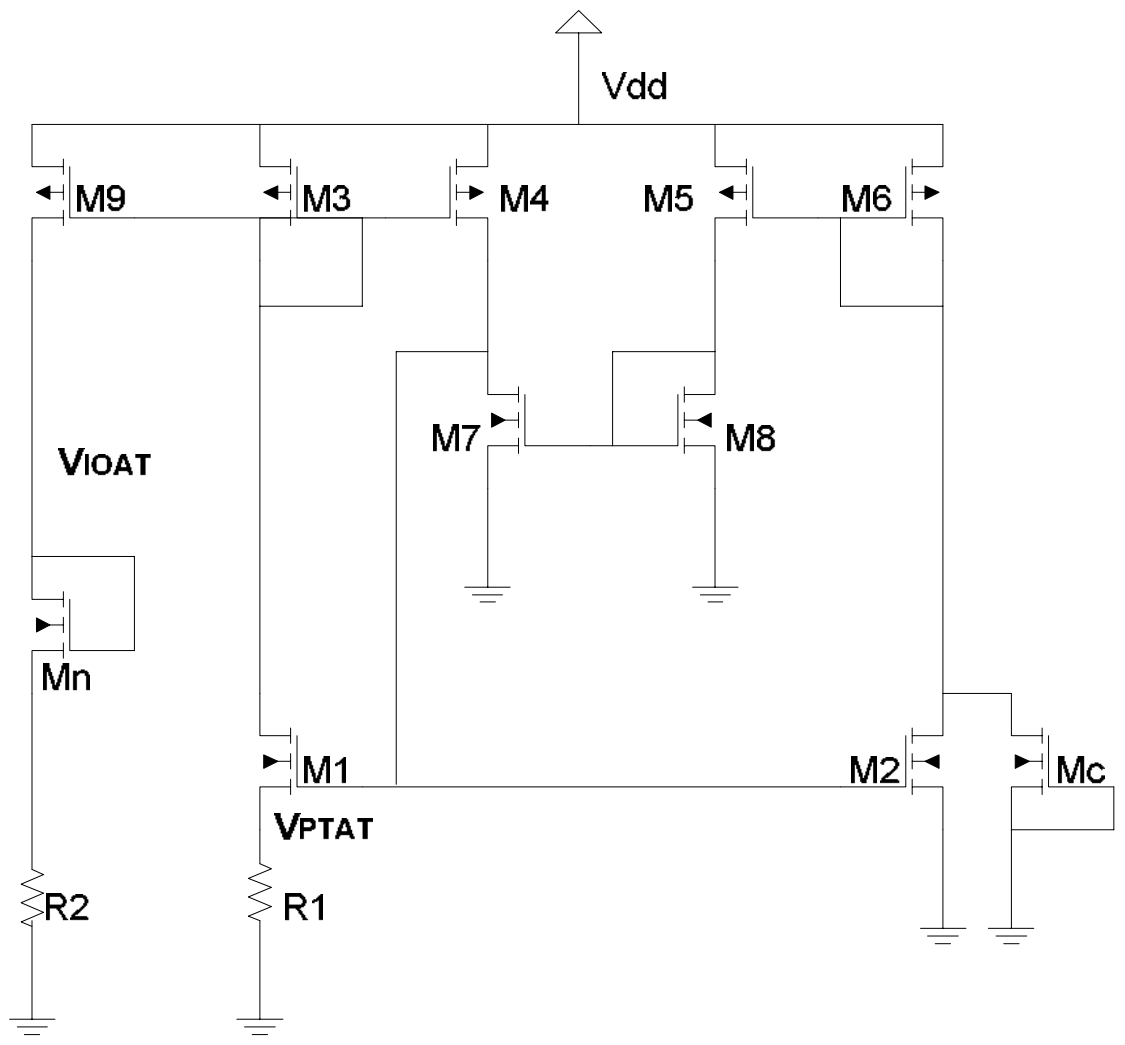

Fig. 4. Resistor-based CMOS PTAT & IOAT references.

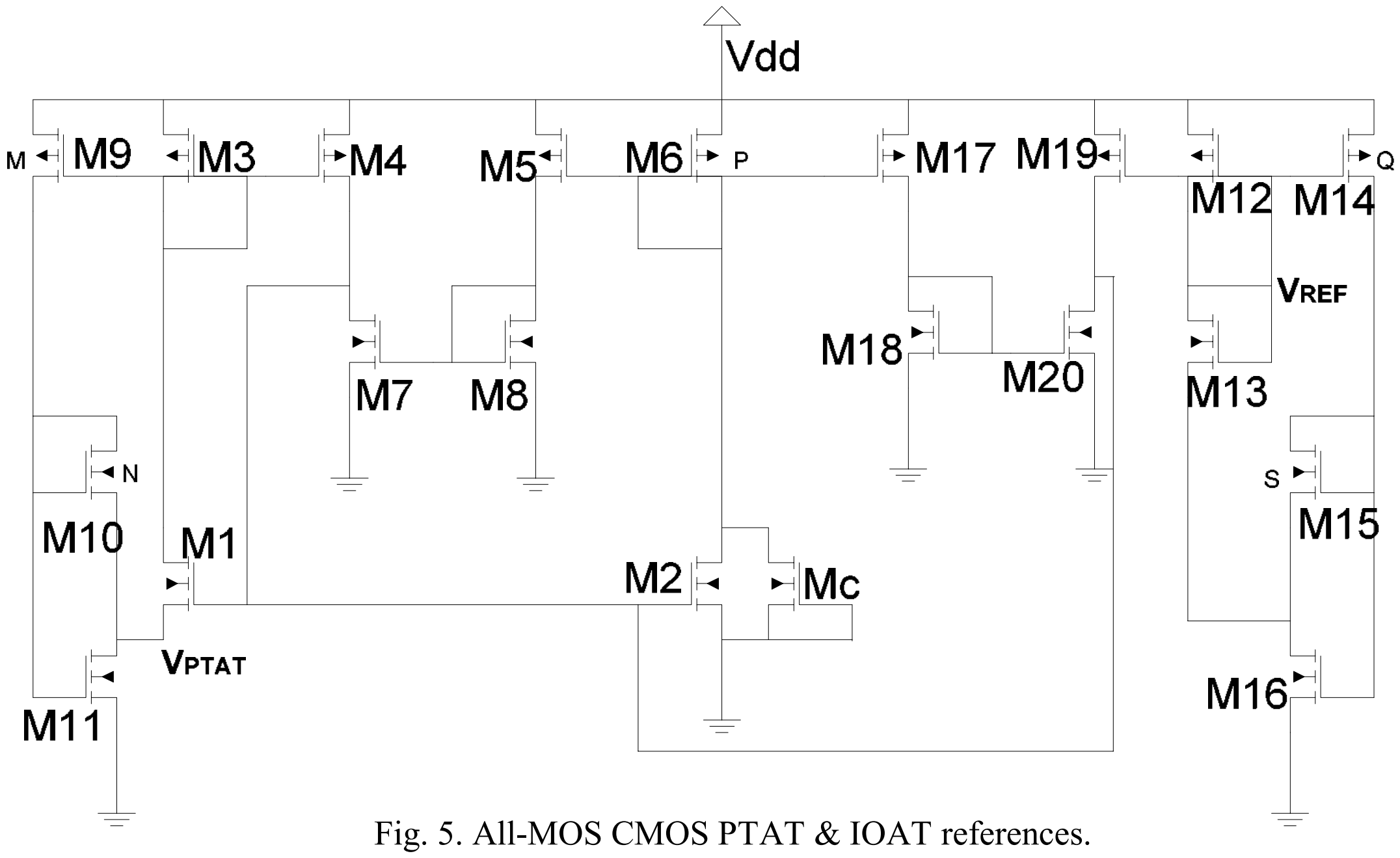

Fig. 5. All-MOS CMOS PTAT & IOAT references.

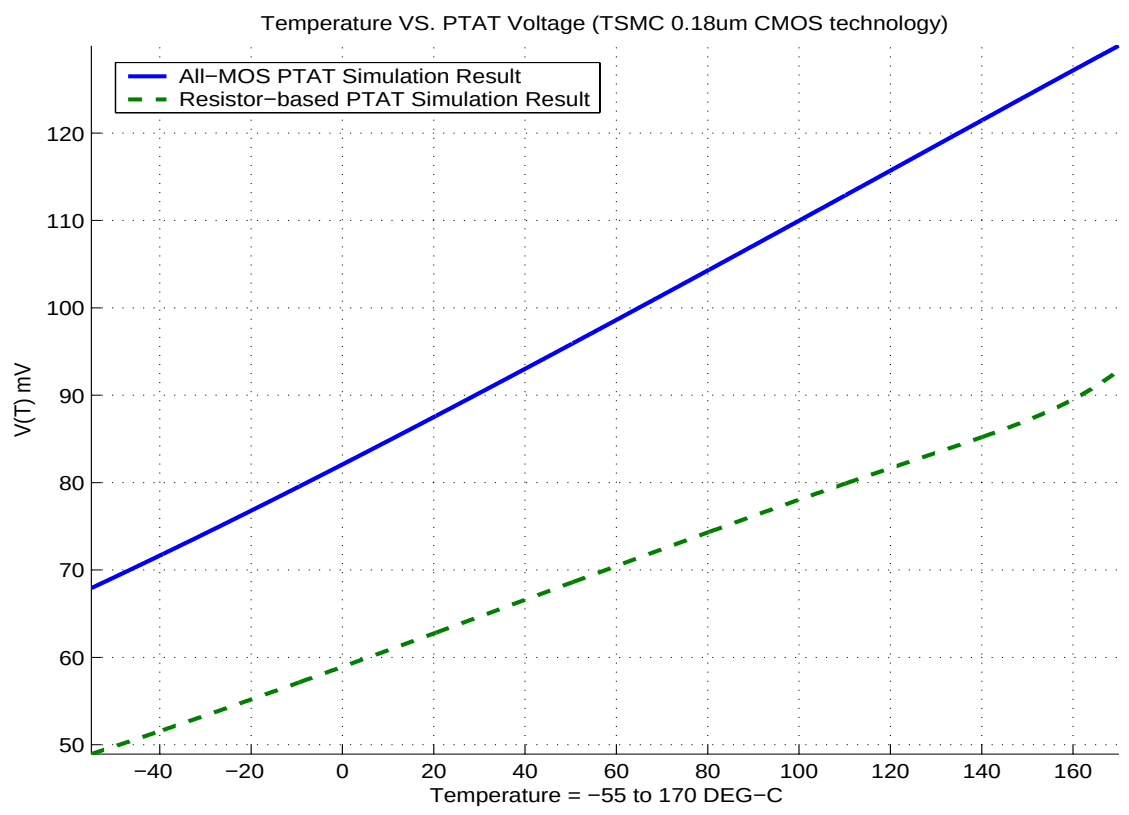

Fig. 6. The simulation result of PTAT references.

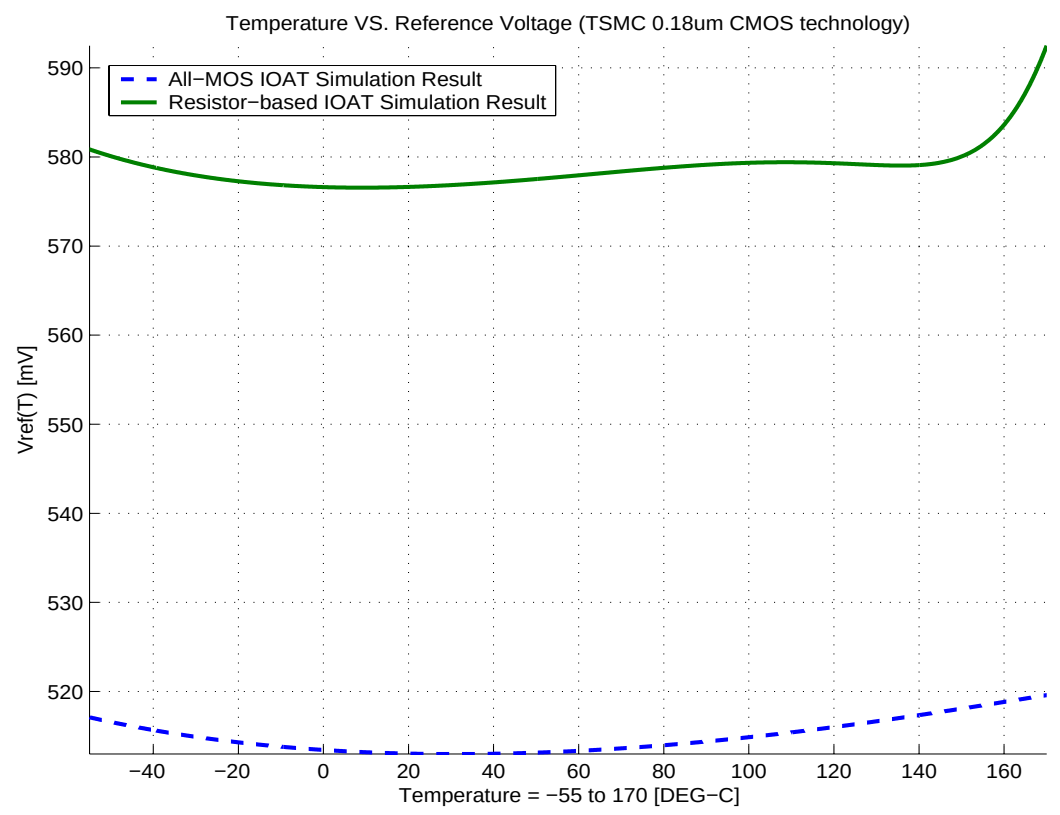

Fig. 7. The simulation result of IOAT references.

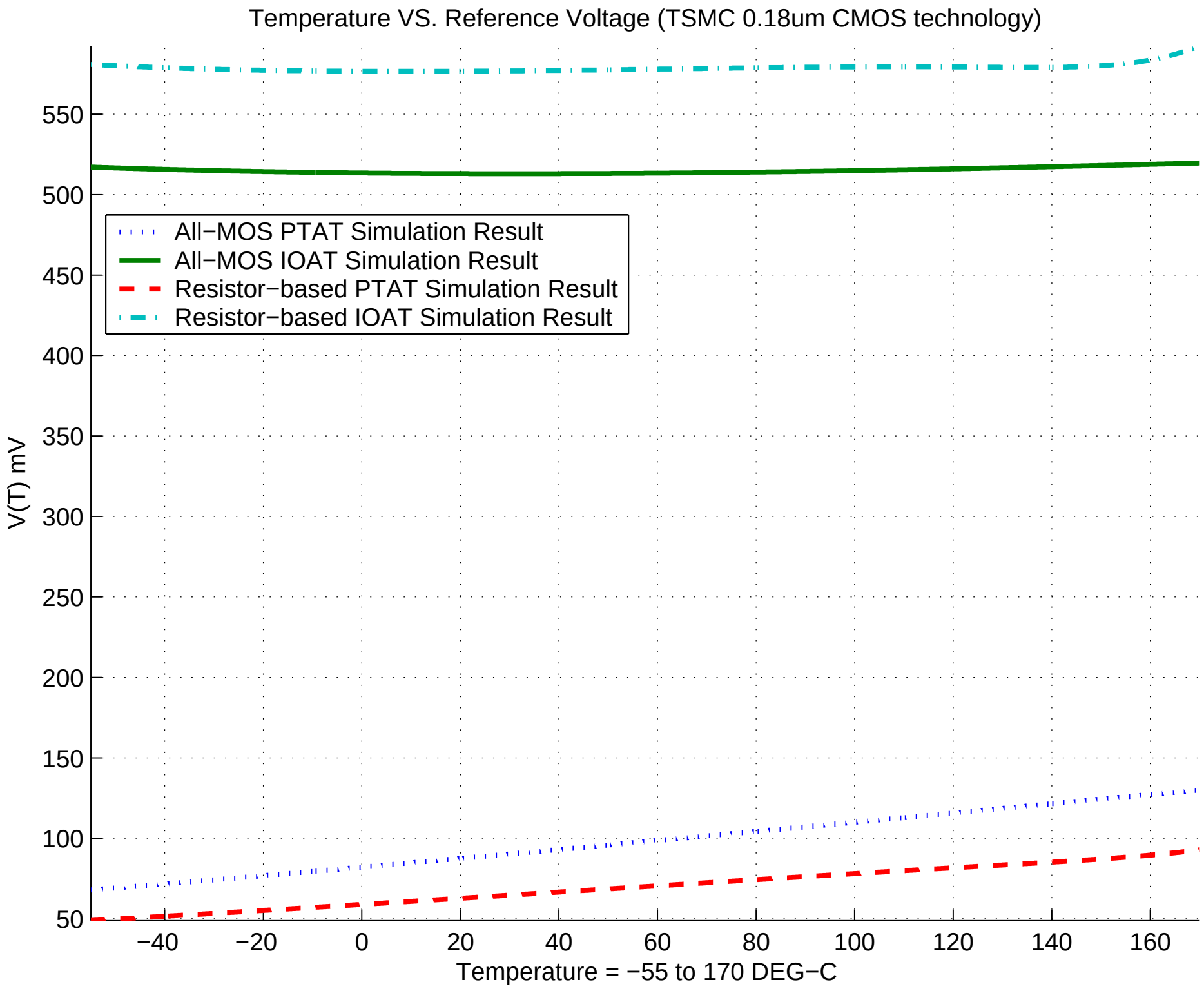

Fig. 8. The simulation result of all voltage references.

## 5. ACKNOWLEDGEMENT

This research was supported by National Science Council, Taiwan (contract number: NSC 95-2220-E-009-029) and Ministry of Education, Taiwan (MoE ATU program.) The Author would like to acknowledge the design parameters provided by National Chip Implementation Center, Taiwan.